\documentclass{aa}
\usepackage{psfig,natbib}
\begin{document}
\headnote{Letter to the Editor}
\title{Direct determination of quasar redshifts}
\titlerunning{Direct quasar redshifts from STJ observations}

\author{%
J.H.J.\ de Bruijne\inst{1},
A.P.\ Reynolds\inst{1},
M.A.C.\ Perryman\inst{1,2},
A.\ Peacock\inst{1},
F.\ Favata\inst{1},
N.\ Rando\inst{1},
D.\ Martin\inst{1},
P.\ Verhoeve\inst{1} \and
N.\ Christlieb\inst{3}}

\authorrunning{J.H.J.\ de Bruijne et al.}

\institute{%
$^{1}$ Astrophysics Division, European Space Agency, ESTEC, Postbus 299, 2200AG Noordwijk, The Netherlands\\
$^{2}$ Sterrewacht Leiden, Postbus 9513, 2300RA Leiden, The Netherlands\\
$^{3}$ Hamburger Sternwarte, Gojenbergsweg 112, 21029 Hamburg, Germany}

\offprints{Jos.de.Bruijne@esa.int}

\date{Submitted to A\&A Letters on October 31, 2001; accepted on November 26, 2001}

\abstract{%
We present observations of 11 quasars, selected in the range $z
\approx 2.2$--$4.1$, obtained with ESA's Superconducting Tunnel
Junction (STJ) camera on the WHT. Using a single template QSO
spectrum, we show that we can determine the redshifts of these objects
to about $1$\%. A follow-up spectroscopic observation of one QSO for
which our best-fit redshift ($z = 2.976$) differs significantly from
the tentative literature value ($z \approx 2.30$) confirms that the
latter was incorrect.
\keywords{%
instrumentation: detectors --
galaxies: distances and redshifts --
galaxies: high-redshift --
quasars: absorption lines --
quasars: emission lines --
quasars: general
}
}

\maketitle

\def\placetableOne{
\begin{table}[t]
\caption[]{The 11 quasars observed. $V$ gives the catalogue magnitude
(probably questionable for 2143$-$158). $T$ gives the exposure time in
seconds, and $z_{\rm obs}$ our estimated redshift. The final two
columns give the literature redshift and its source: S89 =
\citet{ssb89} [spectroscopy]; M77 = \citet{msl77} [objective prism];
C91 = \citet{cfh+91} [spectroscopy]; C85 = \citet{csc85} [grens
plate].}
\renewcommand{\arraystretch}{0.9}
\renewcommand{\tabcolsep}{6.0pt}
\begin{center}
\begin{tabular}{rclrllc}
\noalign{\vskip -0.40truecm}
\noalign{\vskip 0.10truecm}
\hline
\hline
\noalign{\vskip 0.07truecm}
\multicolumn{1}{c}{Obs.} &
\multicolumn{1}{c}{QSO} &
\multicolumn{1}{c}{$V$} &
\multicolumn{1}{c}{$T$} &
\multicolumn{1}{c}{$z_{\rm obs}$} &
\multicolumn{1}{c}{$z_{\rm lit}$} &
\multicolumn{1}{c}{Lit.}\\
\multicolumn{1}{c}{} &
\multicolumn{1}{c}{name} &
\multicolumn{1}{c}{(mag)} &
\multicolumn{1}{c}{(s)} &
\multicolumn{1}{c}{} &
\multicolumn{1}{c}{} &
\multicolumn{1}{c}{$z$}\\
\noalign{\vskip 0.07truecm}
\hline
\noalign{\vskip 0.07truecm}
 1&0000$-$263&17.5 &  600&4.095&4.111&S89\\
 2&0052$-$009&18.2 & 1033&2.190&2.212&C91\\
 3&0055$-$264&17.5 &  600&3.625&3.656&S89\\
 4&0127$+$059&18.0 &  600&2.976&2.30 &M77\\
 5&0132$-$198&18.0 &  900&3.073&3.130&S89\\
 6&0148$-$097&18.4 & 1800&2.845&2.848&S89\\
 7&0153$+$045&18.8 &  600&2.978&2.991&S89\\
 8&0302$-$003&18.4 &  900&3.263&3.286&S89\\
 9&0642$+$449&18.5 &  900&3.366&3.406&S89\\
10&2143$-$158&21.2 & 1800&2.296&2.3  &C85\\
11&2233$+$136&18.6 &  900&3.110&3.209&S89\\
\hline\hline
\end{tabular}
\label{tab:1}
\end{center}
\end{table}
}

\def\placefigureOne{
\begin{figure}[t!]
\centerline{\psfig{file=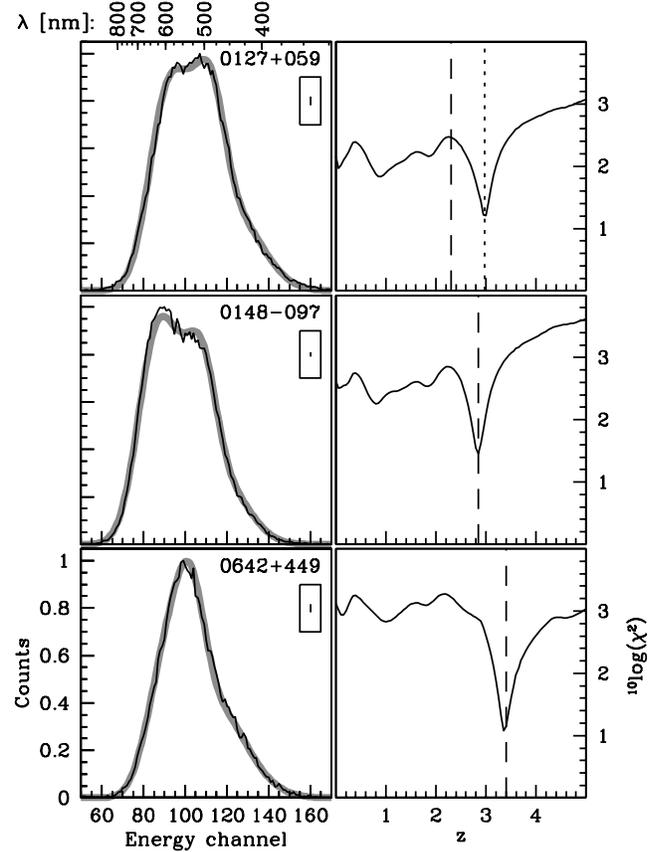,width=8.52truecm,silent=,clip=}}
\caption[]{Results for QSO~0127+059, 0148$-$097, and 0642+449.
{\it Left:\/} the observed (black curves) and modeled (grey curves)
energy channel distributions (arbitrary units). Insets indicate the
estimated Poisson noise. Numbers above the top left panel show the
mapping between energy channel and wavelength. {\it Right:\/} the
corresponding dependence of $\chi^2$ on $z$. Vertical dashed lines
indicate the literature redshifts; the dotted line for QSO~0127+059
indicates the spectroscopic redshift reported in this letter ($z =
3.04$).}
\label{fig:1}
\end{figure}
}

\def\placefigureTwo{
\begin{figure}[t!]
\centerline{\psfig{file=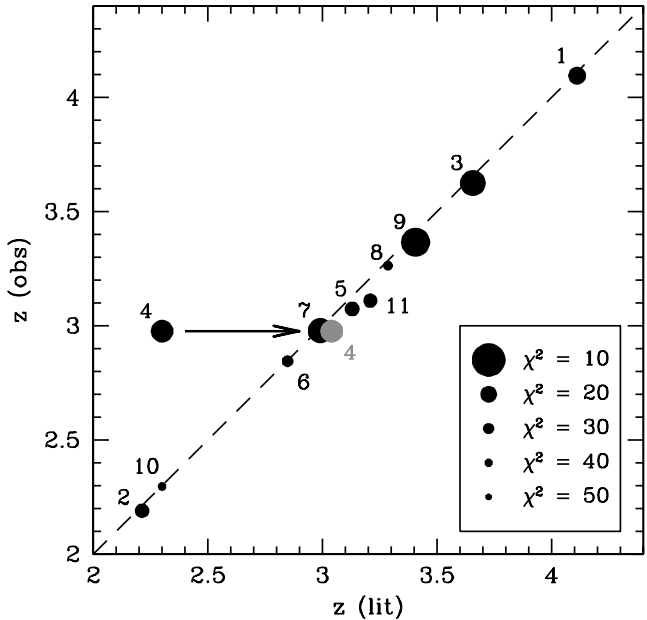,width=8.52truecm,silent=,clip=}}
\caption[]{Observed versus literature redshifts. Numbers refer to the objects 
listed in Table~\ref{tab:1}, and symbol sizes correspond to $\chi^2$
(smaller symbols indicating a poorer fit). QSO~0127+059 has an
incorrect literature redshift of $2.30$; our spectroscopic follow-up
observation yields $z = 3.04$, moving the point to the position shown
in grey. The dashed line shows the nominal 1:1 correlation.}
\label{fig:2}
\end{figure}
}

\def\placefigureThree{
\begin{figure}[t!]
\centerline{\psfig{file=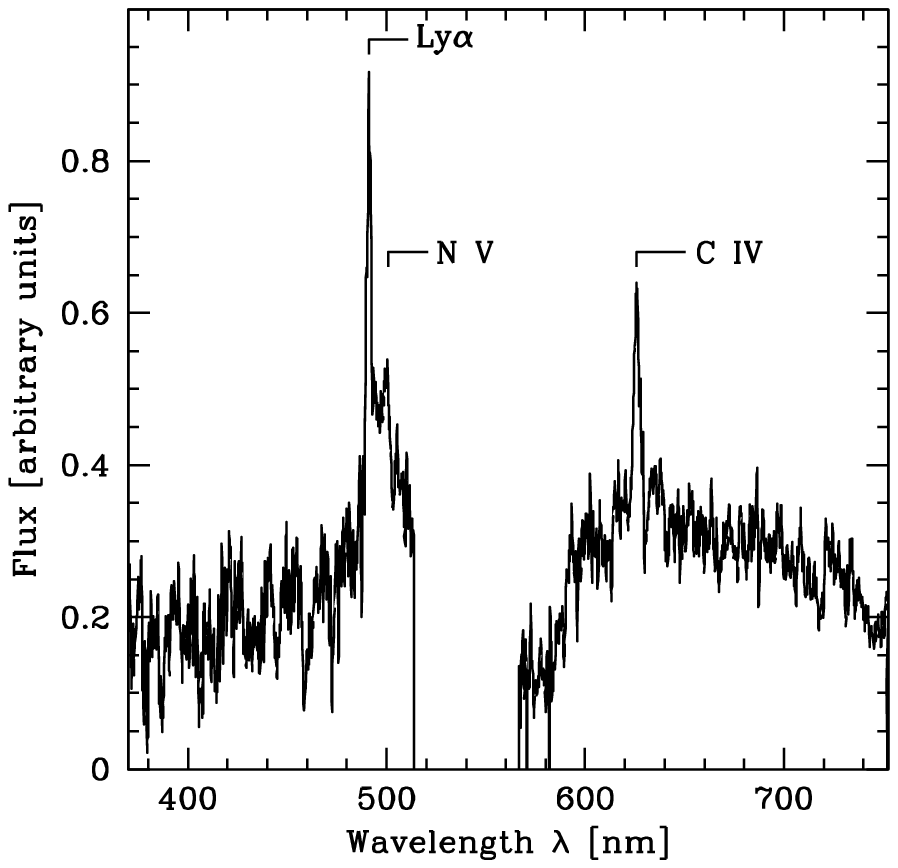,width=8.52truecm,silent=,clip=}}
\caption[]{The spectrum of QSO~0127+059 obtained with the Siding
Spring Observatory $2.3$-m telescope, and smoothed with a $15$ \AA\
FWHM Gaussian. We determine $z = 3.04$; the resulting redshifted
locations of Ly-$\alpha$ (121.6~nm), N{\tt V} (124.0~nm), and C{\tt
IV} (154.9~nm) are indicated.}
\label{fig:3}
\end{figure}
}

\section{Introduction}

Large ground and space telescopes combined with solid state detectors
have revolutionized optical astronomy over the past two decades, yet
deriving physical diagnostics of stars and galaxies still requires the
somewhat indirect methods of filter photometry or dispersive
spectroscopy to measure spectral features, energy distributions, and
redshifts. The recent development of high-efficiency superconducting
detectors \citep{pfp93,pvr+96} has introduced the possibility of
measuring individual optical photon energies directly, and the first
high time-resolution spectrally-resolved observations of rapidly
variable sources such as cataclysmic variables and optical pulsars
using these techniques have been reported
\citep{pfp+99,rmc+99,pcr+01,bcr+01}. Many extensive observational
programmes which aim at determining the large-scale structure of the
Universe, and galaxy formation and evolution (e.g., the Sloan Digital
Sky Survey, \citealt{fss+99}; the Anglo-Australian Telescope 2dF
survey, \citealt{csb+01}), demand high-efficiency extragalactic
spectroscopy. Here we report the first optical measurements of
spectral energy distributions of quasars using an imaging detector
with intrinsic energy resolution, and show that we can determine their
redshifts directly with excellent precision.

\section{Observations}

We observed 11 quasars in the redshift range $z = 2.2$--$4.1$, the
sample comprising relatively bright high-redshift Lyman-limit quasars
from the published literature \citep{ssb89}, supplemented by three
lower redshift objects, two of which were discovered in objective
prism-type surveys (Table~\ref{tab:1}). Observations used the ESA
superconducting tunnel junction (STJ) camera, S-Cam2 \citep{rvv+00},
on the $4.2$-m William Herschel Telescope, La Palma, between 2000
October $1$--$4$. The camera is a $6 \times 6$ array of $25 \times
25$~$\mu$m$^2$ ($0.6 \times 0.6$~arcsec$^2$) tantalum junctions,
providing individual photon arrival time accuracies to about 5~$\mu$s,
a resolving power of ${\cal R} \approx 8$ at $\lambda = 500$~nm, and
high sensitivity from 310~nm (the atmospheric cutoff) to about 720~nm
(currently set by long-wavelength filters to reduce the thermal noise
photons). All objects show strong Ly-$\alpha$ and C{\tt IV} emission
lines which, at these redshifts, will be present within our wavelength
response range. Observations were made in modest seeing (1--1.5~arcsec
at airmass $X = 1$), and at air-masses between $X = 1.07$--$1.82$.

\placetableOne

\section{Data reduction}

Information on each detected photon consists of arrival time, $x,y$
coordinates of the junction, and an energy channel in the range
$0$--$255$. A photon of energy $E_{\rm p}$ (in eV) incident on a
particular junction is assigned to an energy channel $E_i = G \cdot
E_{\rm p} + C$, where each pixel is characterised by its own gain $G$
(in channels per eV) and offset $C$ (in channels). Laboratory
measurements have confirmed that all 36 junctions have a highly
linear, albeit slightly pixel-dependent, energy response. Calibration
consists of first bringing the observed energy channels to a common
reference scale, corresponding to an arbitrary reference pixel, using
a fixed gain map based on laboratory measurements. The offset of the
reference pixel is constant ($C = -2.0$), and its gain is then the
only free parameter in the absolute energy calibration. Small temporal
gain variations resulting from bias voltage drifts and small detector
temperature variations ($\pm$$0.01$~K on the nominal operating
temperature of $\approx$$0.32$~K) are monitored and calibrated.

Subtraction of the appropriate sky contribution for each quasar
spectrum can in principle be based on the background signal in the
outer array junctions, but given the small array size and seeing and
refraction effects, we generally also took a nearby sky frame
immediately following each quasar observation. Most observations were
taken in astronomically dark time; QSO~2233+136, 2143$-$158, and
0148$-$097 were observed with the Moon setting, with background
subtraction slightly less accurate.

\section{Results}

\placefigureOne

We have determined each quasar redshift by comparing the calibrated
energy distributions, $f_{\rm obs}(E_i)$, with a single rest-frame
composite quasar spectrum \citep{zkt+97} based on 284 Hubble Space
Telescope Faint Object Spectrograph spectra of 101 quasars with $z >
0.33$. For a given gain $G$ and redshift $z$, we construct the model
energy-channel distribution $f_{\rm mod}(E_i)$, as follows. The
template spectrum is shifted from the rest frame to redshift $z$, and
a mean accumulated absorption of the Lyman forest for this redshift is
introduced \citep{mj90} (all our objects are at high Galactic
latitude, and we neglect Galactic reddening). The resulting spectrum
is corrected for the mean atmospheric transmission at the relevant
airmass, adjusted for the instrument and telescope efficiency curves
and exposure time, transformed from wavelength spectra to
energy-channel spectra, and finally convolved with a suitable Gaussian
in order to account for the finite energy resolution of the
detector. We then derive redshift $z$ and gain $G$ (and a
normalization constant $N$), by minimizing the classical $\chi^2$
function:
\begin{equation}
\chi^2(z,G,N) = {{1}\over{118}} \cdot \sum_{i=50}^{170}\left[{{f_{\rm obs}(E_i)-N \cdot f_{\rm mod}(E_i)}\over{\sigma_{f_{\rm obs}(E_i)}}}\right]^2,
\end{equation}
using a downhill simplex routine \citep{ptv+95}. Summation extends
over relevant energy channels, and $118$ is the number of degrees of
freedom. Resulting gains are in the range $42$--$45$
channels~eV$^{-1}$, consistent with laboratory calibration. The
resulting redshifts are listed in Table~\ref{tab:1}, along with the
literature values. Examples of the observed and modeled spectra are
shown in Figure~\ref{fig:1}. The overall shape of these spectra, and
in particular the sharp falloff at low energy channels, is due to the
response of the instrument and telescope. In practice, the Ly-$\alpha$
emission line and the associated break at shorter wavelengths
contribute most to the redshift determination. Figure~\ref{fig:2}
compares the best-fit redshifts with the literature values.

QSO~0127+059 is our single prominent outlier. It was discovered in a
thin prism survey \citep{msl77}, classified as a possible quasar, and
tentatively assigned a redshift of $z \approx 2.30$, but with an
uncertain line identification. Although the quality of our fit is
acceptable (Figure~\ref{fig:1}), our derived redshift, $z = 2.976$,
differs significantly from the literature value. We subsequently
obtained a 1200~s spectrum of QSO~0127+059 (Figure~\ref{fig:3}) with
the Siding Spring Observatory $2.3$-m telescope. The wavelength
coverage (not optimised for quasar spectroscopy) was $345$--$537$ and
$560$--$753$~nm, using the Double Beam Spectrograph with dichroic~1
and gratings $600$B and $600$R. We determine a spectroscopic redshift
$z = 3.04$, which agrees with our estimate to about 2\%
(Figure~\ref{fig:2}).

\placefigureTwo

\placefigureThree

A small systematic offset in the overall correlation, of
$\approx$$0.03$ in $z$, can be attributed to a small mismatch in the
shape of the energy broadening function or the overall throughput used
in the modeling. The mean scatter for all observations is $\sigma_z =
0.03$; removing the systematic offset, 8 of the 11 objects agree to
within 1\%. Several factors, such as gain variations, erroneous sky
subtraction or extinction correction (e.g., due to unmodeled seasonal
Saharan dust in the atmosphere), or template mismatch at the object
level (related to continuum slope, line ratios, etc.), may explain the
observed spread. Formally, none of the fits is particularly good, in
the sense that none of them has reduced $\chi^2 \approx 1$. A key
factor in $\chi^2$-statistics, however, is the absence of systematic
errors, which will exist here in part due to template mismatch,
although details are largely hidden as a result of the limited
detector resolution. The general consistency between the models and
the observations, and the pronounced, deep and narrow, minima in all
$\chi^2$ versus $z$ plots, nonetheless indicate that our fits, as a
set, are acceptable.

The pronounced minima are apparent in our data sets truncated {\it a
posteriori} to observation times as small as, e.g. $10$--$20$~s for
the $z = 4.1$ object QSO~0000$-$263, where $\approx$$350$ source
photons s$^{-1}$ were recorded.

\section{Discussion}

Although extraction of detailed physical information from the spectra
is limited by the modest resolving power (${\cal R} \approx 8$) of the
current device, a significant improvement in energy resolution can be
expected in the future \citep{pfp93,pvr+97}, and additional template
spectra could then be used for model fitting. Our results show that
low-resolution spectroscopy of faint extragalactic sources is possible
with these devices, enabling the determination of redshift, and
perhaps morphological type and emission and absorption line ratios
\citep{jak99,mb00x}. Our instrument development is aimed at larger
format arrays to facilitate sky subtraction and possibly for
multi-object spectroscopy, and an increased energy resolution to
improve physical diagnostic capability. An overall wavelength response
extending further to the red, consistent with the fundamental device
response characteristics, would also open up a larger accessible
redshift range.

\begin{acknowledgements}

The William Herschel Telescope is operated on the island of La Palma
by the Isaac Newton Group (ING) in the Spanish Observatorio del Roque
de los Muchachos of the Instituto de Astrof{\'{\i}}sica de
Canarias. We thank J.~Verveer and S.~Andersson for instrument
contributions, P.~Jakobsen for advice on the template spectrum,
I.~Busa and B.~Fuhrmeister for obtaining and reducing the spectrum of
QSO~0127+059, and the referee, Scott Croom, for helpful comments. This
research has made use of the ADS (NASA) and SIMBAD (CDS) services.

\end{acknowledgements}

\end{document}